\documentclass{article}
\usepackage{times}
\begin{document}
\title{Duality, Quantum Mechanics\\ and (Almost) Complex Manifolds}
\author{Jos\'e M. Isidro \\ 
Instituto de F\'{\i}sica Corpuscular (CSIC--UVEG)\\
Apartado de Correos 22085, 46071 Valencia, Spain\\
{\tt jmisidro@ific.uv.es}}
\maketitle

\begin{abstract}
The classical mechanics of a finite number of degrees of freedom requires a symplectic 
structure on phase space ${\cal C}$, but it is independent of any complex structure. 
On the contrary, the quantum theory is intimately linked with the choice 
of a complex structure on ${\cal C}$. When the latter is a complex--analytic manifold 
admitting just one complex structure, there is a unique quantisation whose classical 
limit is ${\cal C}$. Then the notion of coherence is the same for all observers.
However, when ${\cal C}$ admits two or more nonbiholomorphic complex structures, 
there is one different quantisation per different complex structure on ${\cal C}$. 
The lack of analyticity in transforming between nonbiholomorphic complex structures 
can be interpreted as the loss of quantum--mechanical coherence under the corresponding 
transformation. Observers using one complex structure perceive as coherent 
the states that other observers, using a different complex structure, 
do not perceive as such. This is the notion of a quantum--mechanical duality 
transformation: the relativity of the notion of a quantum.


2001 PACS codes: 03.65.Bz, 03.65.-w, 03.65.Ca, 03.65.Sq.

\end{abstract}

\tableofcontents

\section{Introduction}\label{intro}

\subsection{Notations}\label{outline}

Throughout this article, ${\cal C}$ will denote a real $2n$--dimensional symplectic manifold 
({\it classical phase space}\/), endowed with a symplectic form $\omega$
and local Darboux coordinates $q^l$, $p_l$, $l=1,\ldots, n$. A complex structure on ${\cal C}$,
compatible with the symplectic structure, will be denoted by ${\cal J}$, and an almost 
complex structure by $J$. Upon quantisation there will be a complex, separable Hilbert space 
${\cal H}$ of quantum states. The notations 
${\cal F}$ and ${\cal L}$ respectively stand for a foliation of ${\cal C}$ and its leaves.
Holomorphic line bundles over ${\cal C}$, denoted $N({\cal C})$, are classified by the Picard group 
of ${\cal C}$, denoted ${\rm Pic}\,({\cal C})$. The moduli space of complex structures on 
${\cal C}$ that are compatible with $\omega$ will be ${\cal M}({\cal C})$. The group of symplectic 
diffeomorphisms of ${\cal C}$ with respect to $\omega$ ({\it canonical transformations}\/) 
will be denoted ${\rm Sp}({\cal C}, \omega)$, and the group of biholomorphic diffeomorphisms 
of ${\cal C}$ with respect to ${\cal J}$ by ${\rm Diff}({\cal C}, {\cal 
J})$ \cite{ARNOLD, KN, LIBSCHLICHENMAIER} .

\subsection{Motivation}\label{moti}

Duality plays a pivotal role in recent breakthroughs in the quantum theories of fields, 
strings and branes \cite{VAFA}. Under {\it duality}\/ one understands a transformation 
of a given theory, in a certain regime of the variables and parameters that define it, 
into a physically equivalent theory with different variables and parameters. 
Often, what appears to be a highly nontrivial quantum excitation in a given theory 
turns out to be a simple perturbative correction from the viewpoint 
of a theory dual to the original one. This suggests that what constitutes 
a quantum correction may be a matter of convention: the notion of classical 
{\it vs.} quantum is relative to which theory the measurement is made 
from, and that there is a relativity in the notion of a quantum \cite{VAFA}.  
As a consequence, that quantum--mechanical coherence is also theory--dependent, 
or observer--depedent. What one observer calls a {\it coherent state}\/ need not 
be coherent to another observer.

On the other hand, coherent states are quantum--mechanical states that enjoy semiclassical 
properties \cite{PERELOMOV}. One definition of coherent states uses the Heisenberg 
inequality $\Delta q\Delta p \geq \hbar/2$. The latter is saturated precisely 
by coherent states. Planck's constant $\hbar$ can be interpreted formally
as a parameter measuring how far quantum mechanics deviates from classical mechanics
\cite{MATONE}. 

The standard formulation of quantum mechanics does not allow for the relativity 
of the concept of a quantum that underlies the notion of duality.
The limit $\hbar\to 0$ is the {\it  semiclassical regime}, and the limit $\hbar\to\infty$  
is the {\it  strong quantum regime}. Under the usual formulation of quantum mechanics, 
if one observer calls a certain phenomenon {\it semiclassical}, 
then so will it be for all other observers. If one observer calls 
a certain phenomenon {\it strong quantum}, then so will it be for all other observers.
Following the suggestion of ref. \cite{VAFA}, the purpose of this article is to expand on previous 
work \cite{ME} aimed at developing a formalism for quantum--mechanical dualities.

\subsection{Summary of main results}\label{suma}

Our analysis concerns three possibilities for ${\cal C}$: 
the case when it admits no complex structure (compatible with the 
symplectic structure), the case when it admits a unique complex structure
(compatible with the symplectic structure), and the case of more than one complex structure 
(compatible with the symplectic structure). In terms of the moduli space of complex structures, 
these respectively correspond to the cases when ${\cal M}({\cal C})$ is empty, when it contains 
just one point, or more than one point.

Coherent states can always be defined locally, {\it i.e.}, in the neighbourhood of any point  
on ${\cal C}$. This is merely a restatement, in physical terms, of Darboux's theorem for symplectic 
manifolds. In the presence of a complex structure, coherence becomes a global property 
on ${\cal C}$. In the absence of a complex structure, however, the best we can do 
is to combine Darboux coordinates $q^l$, $p_l$ as $q^l+{\rm i}p_l$. Technically this only defines 
an almost complex structure on ${\cal C}$ \cite{KN}. Since the combination $q^l+{\rm i}p_l$ 
falls short of defining a complex structure, quantities depending on $q^l+{\rm i}p_l$ 
on a certain coordinate chart will generally also depend on $q^l-{\rm i}p_l$ 
when transformed to another coordinate chart. This proves that coherence remains 
a local property on classical phase space: observers not connected by means of 
a holomorphic change of coordinates need not, and in general will not,
agree on what is a semiclassical effect {\it vs.} what is a strong quantum effect.

A complex structure on ${\cal C}$ is equivalent to a set of creation and annihilation 
operators for the quantum theory. We have shown in ref. \cite{ME} how ${\cal J}$ gives rise 
to a set of creation and annihilation operators $A_l^{\dagger}$ and $A^l$ on ${\cal H}$, 
and conversely. However, there is one more piece of information that enters the quantum theory.
One has to specify a vacuum state with respect to which the quanta created 
and destroyed by $A^{\dagger}_l$ and $A^l$ are measured. Like the creation and annihilation 
operators, this additional piece of information also has a geometric origin.
It is the Picard group ${\rm Pic}\,({\cal C})$, which parametrises holomorphic 
equivalence classes of holomorphic line bundles $N({\cal C})$ 
\cite{LIBSCHLICHENMAIER}. 
Every element of ${\rm Pic}\, ({\cal C})$ corresponds to a different 
equivalence class. The identity corresponds to the trivial 
line bundle, while elements different from the identity correspond to nontrivial line bundles. 
In ref. \cite{NOS} we have proved that the vacuum state $|0\rangle$ of the quantum 
theory is the fibrewise generator of such line bundles. (Using the 
terminology of ref. \cite{NOS}, in the present article we restrict our 
attention to a given local trivialisation of the quantum Hilbert--space vector 
bundle; we also consider the case of a nondegenerate vacuum. These 
assumptions simplify our analysis without losing generality).

When ${\cal C}$ admits just one complex structure, there is a unique quantum mechanics whose 
classical limit is ${\cal C}$. Then the notion of quantum--mechanical coherence 
is the same for all observers, and there is no room for an observer--dependence 
of the concept of a quantum.  We would like to stress the fact that such 
is the case in all standard applications of quantum mechanics (harmonic oscillator, 
Coulomb potential, angular momentum). This is important, since whatever 
the extension is of quantum mechanics that accommodates dualities, 
it must respect the fact that ordinary quantum mechanics exhibits no 
dualities at all. 

This picture changes when ${\cal M}({\cal C})$ has more than one point, 
{\it i.e.}, when ${\cal C}$  admits two or more nonbiholomorphic complex structures
\cite{LIBSCHLICHENMAIER}. 
Then there is more than one way to define quantum--mechanical creation and annihilation 
operators. Coherent states being defined as eigenstates of the annihilation operator,
there is one different coherent--state quantisation per different complex structure 
on ${\cal C}$. The lack of analyticity in transforming between nonbiholomorphic 
complex structures can be recast as the loss of quantum--mechanical coherence under 
the corresponding transformation. Observers using one complex structure perceive 
as coherent the same states that observers using a different complex structure 
do not perceive as such. This is precisely the notion of duality.

\section{Coherent states on complex phase spaces}\label{globcoh}

The material of this section is well known, but we summarise it here for later use. 
We begin by observing that M--theory dualities are often tested using BPS spectra 
\cite{VAFA}. Such states are stable thanks to supersymmetry. In quantum mechanics, 
the analogues of BPS states are coherent states (which we use here under 
Weyl's presentation \cite{PERELOMOV}). They are also stable under time evolution. 
Thus coherent states are useful kinematical tools to analyse dualities 
in quantum mechanics.

Let us assume that ${\cal C}$ admits a {\it unique}\/ complex structure ${\cal J}$. 
Furthermore let ${\cal J}$ be compatible with the symplectic structure 
$\omega$. This means that the real and imaginary parts of the holomorphic 
coordinates $z^l$ for ${\cal J}$ are Darboux coordinates for the symplectic form 
$\omega$:
\begin{equation}
z^l=q^l + {\rm i} p_l,  \qquad l=1,\ldots, n.
\label{compcoords}
\end{equation}
The set of all $z^l$ so defined provides a holomorphic atlas for ${\cal C}$.
Upon quantisation, the Darboux coordinates
$q^l$ and $p_l$ become operators $Q^l$ and $P_l$ on ${\cal H}$ satisfying the Heisenberg algebra 
\begin{equation}
[Q^j, P_k]={\rm i}\delta^j_k.
\label{hei}
\end{equation}
Define creation and annihilation operators
\begin{equation}
A_l ^{\dagger}= Q^l-{\rm i} P_l,\qquad
A^l=Q^l+{\rm i} P_l,\qquad l=1,\ldots, n.
\label{anni}
\end{equation}
Quantum excitations are measured with respect to a vacuum state 
$|0\rangle$. The latter is defined as that state in ${\cal H}$ which
satisfies
\begin{equation}
A^l|0\rangle = 0,  \qquad l=1, \ldots, n,
\label{vac}
\end{equation}
and coherent states $|z^l\rangle$ are eigenvectors of $A^l$, 
with eigenvalues given in equation (\ref{compcoords}):
\begin{equation}
A^l|z^l\rangle=z^l|z^l\rangle,\qquad l=1,\ldots, n.
\label{annop}
\end{equation}

How do the vacuum state $|0\rangle$ and the coherent states $|z^l\rangle$
transform under a canonical coordinate transformation on ${\cal C}$? 
Call the new Darboux coordinates $q'^l$, $p'_l$. Upon quantisation the 
corresponding operators $Q'^l$, $P'_l$ continue to satisfy the Heisenberg 
algebra (\ref{hei}). Then the combinations
\begin{equation}
z'^l=q'^l + {\rm i} p'_l,  \qquad l=1,\ldots, n
\label{xcompcoords}
\end{equation}
continue to provide holomorphic coordinates for ${\cal C}$, and the 
transformation between the $z^l$ and the $z'^l$ is given by an $n$--variable
holomorphic function $f$,
\begin{equation}
z'=f(z),\qquad \bar\partial f=0.
\label{fhol}
\end{equation}
We can write as above
\begin{equation}
A'^l=Q'^l+{\rm i} P'_l,\qquad l=1,\ldots, n,
\label{annix}
\end{equation}
\begin{equation}
A'^l|0\rangle = 0,  \qquad l=1, \ldots, n,
\label{vacx}
\end{equation}
\begin{equation}
A'^l|z'^l\rangle=z'^l|z'^l\rangle,\qquad l=1,\ldots, n.
\label{annopx}
\end{equation}
There is no physical difference between equations (\ref{anni}), (\ref{vac}) and 
(\ref{annop}), on the one hand, and their holomorphic transforms (\ref{annix}), 
(\ref{vacx}) and (\ref{annopx}), on the other. Under the transformation 
(\ref{fhol}), the vacuum state $|0\rangle$ is mapped into itself, 
and the coherent states $|z^l\rangle$ are mapped into the coherent states $|z'^l\rangle$.
Therefore the notion of coherence is global for all observers on ${\cal 
C}$, {\it i.e.}, any two observers will agree on what is a coherent 
state {\it vs.} what is a noncoherent state. A consequence of this fact is the following. 
Under holomorphic diffeomorphisms of ${\cal C}$, the semiclassical regime of the quantum 
theory is mapped into the semiclassical regime, and the strong quantum regime is mapped 
into the strong quantum regime.

Conversely, one can reverse the order of arguments in this section. 
Start from the assumption that one can define global coherent states 
$|z^l\rangle$ and a global vacuum $|0\rangle$ on the symplectic manifold ${\cal C}$.
{\it Globality}\/ here does not mean that one can cover all of ${\cal C}$ 
with just one coordinate chart (which is impossible if ${\cal C}$ is compact).
Rather it means that, under all
symplectomorphisms of ${\cal C}$, the vacuum is mapped into itself, 
and coherent states are always mapped into coherent states. 
Then the coordinates $z^l$ defined by the eigenvalue equations 
(\ref{annop}) provide a local chart for ${\cal C}$. Collecting together the set of all 
such possible local charts we obtain an atlas for ${\cal C}$. This atlas is holomorphic 
thanks to the property of globality.

To summarise, the existence of a complex structure ${\cal J}$ 
is equivalent to the existence of a globally defined vacuum and globally 
defined coherent states.

\section{Varying the complex structure}\label{qqmod}

Up to now we have assumed that ${\cal C}$ admits a unique 
complex structure ${\cal J}$ which is kept fixed throughout. Next we want 
to study the dependence of the quantum theory on the choice made for 
${\cal J}$, in those cases when ${\cal C}$ admits more than one complex 
structure, compatible with the symplectic structure. Let us first present 
some examples.

The complex 1--dimensional torus $T^2$ is the quotient 
of the complex plane ${\bf C}$, with coordinate $z$, by the action of the group 
${\bf Z}\times {\bf Z}$ whose generators are $1$ and $\tau$, where ${\rm Im}\,\tau>0$:
\begin{equation}
z\simeq z+1, \qquad z\simeq z+\tau.
\label{eqq}
\end{equation}
The moduli space ${\cal M}(T^2)$ is the quotient of the upper half--plane
by the action of ${\rm SL}(2, {\bf Z})$ \cite{LIBSCHLICHENMAIER}.
In geometric quantisation $T^2$ is quantised as follows \cite{WOODHOUSE}.
The K\"ahler form is  
\begin{equation}
\omega={{\rm i}\pi\over {\rm Im}\, \tau}\, {\rm d}z\wedge{\rm d}{\bar z}.
\label{ffmm}
\end{equation}
The quantum line bundle is the theta line bundle of degree 1, {\it i.e.}, 
the holomorphic line bundle whose global holomorphic sections are multiples of the Riemann 
theta function 
\begin{equation}
\theta(z, \tau)=\sum_{n\in{\bf Z}}{\rm exp}\,\left({\rm i}\pi \tau n^2 + 
2 {\rm i}\pi nz\right).
\label{teta}
\end{equation} 
This wavefunction generates the Hilbert space ${\cal H}$. We see that the quantum 
theory depends explicitly on the complex structure $\tau$. 

Let us now generalise the previous conclusions following a path--integral approach.
The quantum--mechanical transition amplitude from the initial state 
$|q_1,t_1\rangle$ to the final state $|q_2, t_2\rangle$ is given by the 
phase--space path--integral
\begin{equation}
\langle q_2, t_2|q_1, t_1\rangle = 
\int {\rm D}p {\rm D}q \,{\rm exp}\left({\rm i}\int_{t_1}^{t_2}{\rm d}t
\left( p\dot q - H(p,q)\right)\right).
\label{ampfey}
\end{equation}
When $H$ depends quadratically on $p$ two simplifications occur. First, we 
have $p\dot q - H=L$. Second, the integral over the momentum is Gaussian and
can be carried out explicitly. This gives, up to normalisation, 
\begin{equation}
\langle q_2, t_2|q_1, t_1\rangle =
\int {\rm D} q\, {\rm exp}\left({\rm i} \int_{t_1}^{t_2}{\rm d} t \,L(\dot 
q, q)\right).
\label{empfey}
\end{equation}
In a Hamiltonian approach, amplitudes are computed according to eqn. (\ref{ampfey}). 
Then the path integral extends to all configurations of coordinate and momentum that 
are compatible with a given complex structure on the torus. That is, a given complex 
structure must be specified first and kept fixed throughout. Complex structures are not 
integrated over in the Feynman path integral (\ref{ampfey}). These facts have a counterpart 
in the Lagrangian approach (\ref{empfey}), where one integrates over $q$ only.

\section{Quantum mechanics on complex phase spaces}\label{hache}

We are now ready to generalise the previous conclusions to an arbitrary complex phase space 
${\cal C}$. The latter carries a symplectic structure compatible with a complex structure.
For an introduction to the theory of deformations of complex 
structures, see ref. \cite{BRUZZO}.

\subsection{Creation and annihilation operators: the choice of a complex structure}\label{buono}

Upon quantisation, the local Darboux coordinates $q^l$ and $p_l$ become selfadjoint 
operators $Q^l$ and $P_l$ on ${\cal H}$ satisfying the Heisenberg algebra.
One defines the creation and annihilation operators as in eqn. (\ref{anni}).
Then coherent states $|z^l\rangle$ are given as in eqn. (\ref{annop}).
While the classical theory made use of coordinates 
and momenta, the quantum theory naturally combines them into the complex combinations 
(\ref{anni}). This natural pairing of $q^l$ and $p_l$ was absent 
in classical mechanics, where the complex combination $z^l=q^l + {\rm i} p_l$ 
was at most a useful artifact, devoid of any physical meaning. In the 
quantum theory, the choice of a set of creation and annihilation operators 
is equivalent to the choice of a complex structure on ${\cal C}$.

\subsection{Nonbiholomorphic complex structures: nonequivalent quantisations}\label{vvar}

A complex structure is a {\it global}\/ choice, modulo biholomorphic diffeomorphisms of ${\cal C}$, 
of a set of creation and annihilation operators for a quantum mechanics. We will see in section 
\ref{loccoh} that an almost complex structure is a {\it local}\/ choice of a set of creation and 
annihilation operators, that cannot be extended globally over all ${\cal C}$.

Quantum mechanics may be understood as performing an infinite expansion 
in powers of $\hbar$ around a certain classical mechanics. 
So above, {\it globality}\/ means that any two observers on ${\cal C}$ agree, 
order by order in this expansion, 
in their respective descriptions of any given quantum phenomenon. 
In particular, if one observer perceives this phenomenon as semiclassical, 
then so will the other observer. If one observer perceives 
it as strong quantum, then so will the other observer. 
As a consequence, globality implies that any two observers will agree 
on the notion of coherence. On the contrary, {\it locality}\/ means disagreement 
between different observers of the same given quantum phenomenon, in their respective 
descriptions as an expansion in powers of $\hbar$. 

We ask ourselves if it is possible to allow for an observer--dependence 
of the vacuum state and of the notions of a quantum and of quantum--mechanical coherence, 
without renouncing the power of complex analysis on ${\cal C}$. We are thus led to considering 
two or more nonbiholomorphic complex structures ${\cal J}^{(\alpha)}$ 
labelled by an index $\alpha$. The latter runs over a certain moduli space 
${\cal M}({\cal C})$.

Since ${\cal C}$ is assumed to be a complex manifold, there is at least 
one point  $\alpha\in {\cal M}({\cal C})$. The corresponding 
complex structure ${\cal J}^{(\alpha)}$ gives rise, in coherent--state 
quantisation, to a globally--defined vacuum state $|0(\alpha)\rangle$, 
plus a family of globally--defined creation and annihilation operators $A^j(\alpha)$, 
$A_k^{\dagger}(\alpha)$, satisfying
\begin{equation}
A^j(\alpha)|0(\alpha)\rangle =0,\qquad 
\left[ A^j(\alpha), A_k^{\dagger}(\alpha)\right]=\delta^j_{k}(\alpha).
\label{unox}
\end{equation}
Coherent states $|z^l(\alpha)\rangle$ are eigenvectors of the annihilation 
operators $A^l(\alpha)$, with eigenvalues $z^l(\alpha)$ that provide holomorphic 
coordinates with respect to the complex structure ${\cal J}^{(\alpha)}$:
\begin{equation}
A^l(\alpha)|z^l(\alpha)\rangle = z^l(\alpha) |z^l(\alpha)\rangle, 
\qquad l=1,\ldots, n.
\label{afa}
\end{equation}
The corresponding quantum mechanics can be entirely expressed in terms of 
these coherent states. Now, if ${\cal M}({\cal C})$ consists of 
just one point, then eqns. (\ref{unox}), (\ref{afa}) are the end of the story, 
and there is no possibility for an observer--dependence of the concept of 
a quantum.

Next assume that there exists a second point $\beta\in {\cal M}({\cal C})$, 
$\beta\neq \alpha$. Eqns. (\ref{unox}), (\ref{afa}) hold 
with $\beta$ replacing $\alpha$. Now the complex structures ${\cal J}^{(\alpha)}$
and ${\cal J}^{(\beta)}$ are nonbiholomorphic. There exists
no biholomorphic diffeomorphism between the $z^l(\alpha)$ 
and the $z^l(\beta)$. We can reexpress this by saying that we have 
two different quantum--mechanical theories, call them QM$(\alpha)$ 
and QM$(\beta)$, both possessing the same classical 
limit ${\cal C}$. Theory QM$(\alpha)$  has the vacuum $|0(\alpha)\rangle$, 
the creators $A_l^{\dagger}(\alpha)$ and the annihilators $A^l(\alpha)$.
Theory QM$(\beta)$ has the vacuum  $|0(\beta)\rangle$, the creators 
$A_l^{\dagger}(\beta)$ and the annihilators $A^l(\beta)$. Since the 
transformation from QM$(\alpha)$ to QM$(\beta)$ is nonholomorphic, 
it follows that the concept of a quantum depends on the observer.
One observer on ${\cal C}$  using theory  QM$(\alpha)$ observes  
this phenomenon as a certain expansion in powers of $\hbar$, another 
observer using theory QM$(\beta)$ does the same. However, there is at 
least one order ${\cal O}(\hbar^m)$ in their respective expansions 
in which the two observers differ---if they did not differ, the transformation between 
$z^l(\alpha)$ and $z^l(\beta)$ would be biholomorphic, contrary to assumption.

The latter statement needs some clarification. On first sight there appears 
to be no link between the lack of analyticity in $z^l$, on the one hand, 
and different power--series expansions in $\hbar$, on the other. That there 
is in fact a link can be seen as follows. Coherent states are quantum--mechanical 
states with semiclassical properties. They saturate Heisenberg's inequality 
$\Delta q\Delta p \geq \hbar/2$. As such they provide the first nontrivial 
order in any series expansion in powers of $\hbar$; corrections to this correspond 
to higher--order effects that can be neglected in a semiclassical approximation. 
The loss of coherence in the transformation between the states $|z^l(\alpha)\rangle$ 
and $|z^l(\beta)\rangle$ means that already at the first nontrivial order the two 
theories QM$(\alpha)$ and QM$(\beta)$ disagree.

\subsection{Dualities on complex phase spaces}\label{ccnn}

A quantum--mechanical duality appears as the possibility of describing 
one given quantum phenomenon in terms of two or more different series 
expansions in powers of $\hbar$. In this section we will put forward 
a proposal for accommodating duality transformations in quantum mechanics
when classical phase space is a complex manifold.

The jacobian matrix of a symplectic transformation has unit determinant.  
Under the assumption of compatibility made in eqn. (\ref{compcoords}), 
we have that symplectic transformations are holomorphic. 
However the converse is not true, as holomorphic transformations need not 
have unit determinant. Compatibility between ${\cal J}$ and $\omega$ implies 
that only holomorphic transformations with unit determinant are symplectic. 
We can rewrite compatibility as the inclusion
\begin{equation}
{\rm Sp}({\cal C}, \omega)\subset {\rm Diff}({\cal C}, {\cal J}).
\label{iinncc}
\end{equation}
Implicitly assumed above is the fact that analyticity refers 
to a fixed complex structure ${\cal J}$. It is also assumed that the 
symplectic property refers to a given symplectic structure $\omega$.
However, assuming the existence of two or more nonbiholomorphic complex structures
on ${\cal C}$, say ${\cal J}^{(\alpha)}$ and ${\cal J}^{(\beta)}$, 
both compatible with $\omega$, we have the inclusions 
\begin{equation}
{\rm Sp}({\cal C, \omega})\subset {\rm Diff}({\cal C}, {\cal J}^{(\alpha)}),
\qquad
{\rm Sp}({\cal C}, \omega)\subset {\rm Diff}({\cal C}, {\cal J}^{(\beta)}).
\label{incuno}
\end{equation}
Symplectomorphisms are holomorphic with respect to all compatible complex structures,
and dualities arise as follows. Assume embedding ${\rm Sp}({\cal C}, \omega)$ 
into two different groups of biholomorphic diffeomorphisms of ${\cal C}$
as in eqn. (\ref{incuno}). Then, as seen from the complex structure ${\cal J}^{(\beta)}$,
dualities are elements of ${\rm Diff}({\cal C}, {\cal J}^{(\alpha)})$ that do 
not belong to ${\rm Sp}({\cal C}, \omega)$. Depending on the topology of 
${\cal C}$, the above expressions may need some natural modifications. For example, when 
${\cal C}$ is compact, only the constant function is everywhere holomorphic.
Then biholomorphic diffeomorphisms are to be considered only on an open neighbourhood, 
instead of over all ${\cal C}$.

In principle, any two points in the moduli space ${\cal M}({\cal C})$ determine 
a duality transformation of the quantum theory. Given ${\cal C}$, an interesting 
question is to identify what regions of ${\cal M}({\cal C})$ are physically accessible. 

\section{Quantum mechanics on almost complex phase spaces}\label{}

\subsection{Coherent states}\label{loccoh}

We now relax the conditions imposed on ${\cal C}$. In this section we will assume
that ${\cal C}$ carries an almost complex structure $J$ compatible with 
the symplectic structure $\omega$. 

Specificallly, an almost complex structure is defined  
as a tensor field $J$ of type $(1,1)$ such that, at every point of 
${\cal C}$, $J^2=-{\bf 1}$ \cite{KN}.  Using Darboux coordinates $q^l$, 
$p_l$ on ${\cal C}$ let us form the combinations
\begin{equation}
w^l=q^l + {\rm i} p_l, \qquad l=1,\ldots, n.
\label{combw}
\end{equation} 
Compatibility between $\omega$ and $J$ means that we can take $J$ 
to be 
\begin{equation}
J\left({\partial\over\partial w^l}\right)={\rm i}{\partial\over\partial w^l},
\qquad 
J\left({\partial\over\partial \bar w^l}\right)=-{\rm i}{\partial\over\partial\bar w^l}.
\label{jota}
\end{equation}
Unless ${\cal C}$ is a complex  manifold to begin with, equations 
(\ref{combw}) and (\ref{jota}) fall short of defining a complex structure ${\cal J}$. 
The set of all such $w^l$ does not provide a holomorphic atlas for ${\cal C}$. 
There exists at least one canonical coordinate transformation between 
Darboux coordinates, call them $(q^l, p_l)$ and $(q'^l, p'_l)$, such that 
the passage between $w^l=q^l+{\rm i}p_l$ and $w'^l=q'^l+{\rm i}p'_l$
is given by a nonholomorphic function $g$ in $n$ variables,
\begin{equation}
w'=g(w,\bar w), \qquad \bar\partial g\neq 0.
\label{nonhol}
\end{equation}

Mathematically, nonholomorphicity implies the mixing of $w^l$ and $\bar w^l$.
Quantum--mechanically, the loss of holomorphicity has deep physical consequences.
One would write, in the initial coordinates $w^l$, a defining equation for the vacuum 
state $|0\rangle$
\begin{equation}
a^l|0\rangle =0, \qquad l=1,\ldots, n,
\label{xvac}
\end{equation}
where $a^l=Q^l + {\rm i} P_l$ is the corresponding local annihilation operator.
However, one is just as well entitled to use the new coordinates $w'^l$
and write
\begin{equation}
a'^l|0'\rangle =0, \qquad l=1,\ldots, n,
\label{xvacu}
\end{equation}
where we have primed the new vacuum, $|0'\rangle$. 
Are we allowed to identify the states $|0\rangle$ and $|0'\rangle$? 
We could identify them if $w'^l$ were a holomorphic function of $w^l$; 
such was the case in section \ref{globcoh}. However, now we are considering 
a nonholomorphic transformation, and we cannot remove the 
prime from the state $|0'\rangle$. This is readily proved.
We have
\begin{equation}
a'=G(a, a^{\dagger}),
\label{effe}
\end{equation}
with $G$ a quantum nonholomorphic function corresponding to the classical 
nonholomorphic function $g$ of equation (\ref{nonhol}). As $[a^j,a_k^{\dagger}]=\delta^j_k$, 
ordering ambiguities will arise in the construction of $G$ from $g$,
that are usually dealt with by normal ordering. Normal ordering would 
appear to allow us to identify the states $|0\rangle$ and $|0'\rangle$.
However this is not the case, as there are choices of $g$ that are left invariant under 
normal ordering, such as the sum of a holomorphic function plus
an antiholomorphic function, $g(w,\bar w) = g_1(w) + g_2(\bar w)$.
Under such a transformation one can see that the state $|0\rangle$ satisfying 
eqn. (\ref{xvac}) will not satisfy eqn. (\ref{xvacu}). 
We conclude that, in the absence of a complex structure 
on classical phase space, the vacuum depends on the observer. 
The state $|0\rangle$ is only defined locally on ${\cal C}$; it cannot be 
extended globally to all of ${\cal C}$. 

Similar conclusions may be expected for the coherent states $|w^l\rangle$. 
The latter are defined only locally, as eigenvectors of the local annihilation 
operator, with eigenvalues given in equation (\ref{combw}):
\begin{equation}
a^l|w^l\rangle=w^l|w^l\rangle,\qquad l=1,\ldots, n.
\label{annopp}
\end{equation}
Due to $[a^j,a_k^{\dagger}]=\delta^j_k$, under the nonholomorphic 
coordinate transformation (\ref{nonhol}), the local coherent states 
$|w^l\rangle$ are {\it not} mapped into the local coherent states 
satisfying
\begin{equation}
a'^l|w'^l\rangle=w'^l|w'^l\rangle,\qquad l=1,\ldots, n
\label{annoppp}
\end{equation}
in the primed coordinates. No such problems arose for the holomorphic operator 
equation $A'=F(A)$ corresponding to the holomorphic coordinate change $z'=f(z)$ 
of equation (\ref{fhol}), because the commutator $[A^j, A_k^{\dagger}]=\delta^j_k$ 
played no role. Thus coherence becomes a local property on classical phase space.
In particular,  observers not connected by means of a holomorphic change 
of coordinates need not, and in general will not, agree on what is a semiclassical 
effect {\it vs.} what is a strong quantum effect. 

As in section \ref{globcoh}, one can reverse the order of arguments. 
Start from the assumption that, around every point on ${\cal C}$, one can define local 
coherent states $|w^l\rangle$ and a local vacuum $|0\rangle$, that however fall short
of being global. This means that there exists at least one symplectomorphism of ${\cal C}$
that does {\it not}\/ preserve the globality property. Local coordinates $w^l$ around 
any point are defined by the eigenvalue equations (\ref{annopp}). Collecting together 
the set of all such possible local charts we obtain an atlas for ${\cal C}$. 
However, unless the local coherent states $|w^l\rangle$ are actually global, 
this atlas is nonholomorphic. This defines an almost complex structure $J$.

To summarise, the existence of an almost complex structure $J$ 
is equivalent to the existence of a locally--defined vacuum and
locally--defined coherent states. When the latter are actually global, 
then $J$ lifts to a complex structure ${\cal J}$, whose 
associated almost complex structure is $J$ itself.

When are local coherent states also global? This question can be recast 
mathematically as follows: when does an almost complex structure $J$ lift 
to a complex structure ${\cal J}$? The almost complex structure $J$ is said 
{\it integrable} when the Lie bracket $[Z, W]$ of any two holomorphic vector 
fields $Z$, $W$ on ${\cal C}$ is holomorphic. A necessary and sufficient condition 
for $J$ to be integrable is the following. Define the tensor field $N$
\begin{equation}
N(Z,W)=\left[Z,W\right] - \left[JZ,JW\right] 
+J\left[Z, JW\right] + J\left[JZ, W\right].
\label{tors}
\end{equation}
Now the almost complex structure $J$ lifts to a complex structure 
${\cal J}$ if and only if the tensor $N$ vanishes identically \cite{KN}.
In the particular case of real dimension 2, every almost complex structure 
is integrable. 
We can turn things around and recast the previous theorem in physical terms: 
when the commutator of any two (infinitesimal) canonical transformations on ${\cal C}$ 
maps coherent states into coherent states, then ${\cal C}$ admits a complex structure. 
The latter is the lift of the almost complex structure $J$ defined 
by $q^l+{\rm i}p_l$ in terms of Darboux coordinates $q^l$, $p_l$. Conversely, 
if a canonical transformation on ${\cal C}$ maps coherent states into 
noncoherent, or viceversa, then $J$ does not lift to a complex 
structure.

\subsection{Proof of coherence}\label{prov}

We have called {\it coherent} the states constructed in previous sections.
However, we have not verified that they actually satisfy the usual requirements 
imposed on coherent states \cite{PERELOMOV}. What ensures that the states so 
constructed are actually coherent is the following argument. 
Any dynamical system with $n$ degrees of freedom that can be transformed 
into a freely--evolving system can be further mapped into the $n$--dimensional harmonic 
oscillator. The combined transformation is canonical. Moreover it is locally 
biholomorphic when ${\cal C}$ is a complex manifold. Thus locally on ${\cal C}$, 
the states $|z^l\rangle$ of section \ref{globcoh} coincide with the coherent states of the 
$n$--dimensional harmonic oscillator. Mathematically this fact reflects the structure 
of a complex manifold: locally it is always biholomorphic with 
(an open subset of) ${\bf C}^n$. Physically this fact reflects 
the decomposition into the creation and annihilation modes of perturbative quantum 
mechanics and field theory. In this way, the mathematical problem of patching together different 
local coordinate charts $(U_{\alpha}, z^l_{\alpha})$ labelled by an index 
$\alpha$ may be recast in physical terms. It is the patching together of different 
local expansions into creators $A^{\dagger}_{\alpha}$ and annihilators $A^{\alpha}$, 
for different values of $\alpha$.

In particular, we can write the resolution of unity on ${\cal H}$ 
associated with a holomorphic atlas on ${\cal C}$ consisting of charts
$(U_{\alpha},z^l_{\alpha})$:
\begin{equation}
\sum_{\alpha}\sum_{l=1}^n\int_{\cal C} {\rm d}\mu_{\cal C}\, |z^l_{\alpha}\rangle  \langle 
z^l_{\alpha}|= {\bf 1},
\label{res}
\end{equation}
where ${\rm d}\mu_{\cal C}$ is an appropriate measure (an 
$(n,n)$--differential) on ${\cal C}$. 

Analogous arguments are also applicable to the states $|w^l\rangle$ 
of section \ref{loccoh}. Every coordinate chart on ${\cal C}$ 
is diffeormorphic to (an open subset of) ${\bf R}^{2n}$, so the $|w^l\rangle$ 
look locally like the coherent states of the $n$--dimensional harmonic oscillator.
However, the loss of holomorphicity of ${\cal C}$ 
alters equation (\ref{res}) in one important way. We may write as above
\begin{equation}
\sum_{\alpha}\sum_{l=1}^n\int_{\cal C} {\rm d}\mu_{\cal C}\, |w^l_{\alpha}\rangle  \langle 
w^l_{\alpha}|,
\label{resx}
\end{equation}
but we can no longer equate this to the identity on 
${\cal H}$. The latter is a {\it complex}\/ vector space, while eqn.
(\ref{resx}) allows one at most to expand an arbitrary,
real--analytic function on ${\cal C}$, since the latter is just a 
real--analytic manifold. Hence we cannot equate (\ref{resx})
to ${\bf 1}_{\cal H}$. We can only equate it to the identity on the 
{\it real}\/
Hilbert space of real--analytic functions on ${\cal C}$.
This situation is not new; coherent states without a resolution of unity have been 
analysed in the literature, where they have been related to the choice of 
an inadmissible fiducial vector. It is tempting to equate this latter 
choice with the viewpoint advocated here about the vacuum state.

\subsection{Dualities on almost complex phase spaces}\label{nonint}

Let us consider the case when ${\cal C}$ admits a certain foliation ${\cal F}$ 
by holomorphic, symplectic leaves ${\cal L}$ \cite{DEWITT}. 
For simplicity we will make a number of technical assumptions. 
First, the leaves ${\cal L}$ have constant real dimension $2m$, 
where $0<2m<2n$; $m$ is called the {\it rank}\/ of the foliation ${\cal F}$. 
We will use the notation $\tilde{\cal L}$ to denote the $2(n-m)$--dimensional 
complement of the ${\cal L}$ in ${\cal C}$. We will assume maximality of 
the rank $m$, {\it i.e.}, no holomorphic leaf exists with dimension 
greater than $2m$. Second, we suppose that the restrictions 
$\omega_{\cal L}$  and $\omega_{\tilde {\cal L}}$ of the symplectic form $\omega$ 
render the leaves ${\cal L}$ and their complements $\tilde{\cal L}$ symplectic.
Third we assume that, on the ${\cal L}$, the complex structure is 
compatible with the symplectic structure as in section \ref{globcoh}. 
Fourth, the complement $\tilde {\cal L}$ is also assumed to carry an almost 
complex structure compatible with $\omega_{\tilde {\cal L}}$ as in section \ref{loccoh}.
All these assumptions amount to a decomposition of $\omega$
as a sum of two terms,  
\begin{equation}
\omega=\omega_{\cal L} + \omega_{\tilde {\cal L}},
\label{twot}
\end{equation}
where in local Darboux coordinates around a basepoint $b\in {\cal C}$ we have
\begin{equation}
\omega_{\cal L}=
\sum_{k=1}^{m} {\rm d}p_k\wedge {\rm d}q^k, \qquad 
\omega_{\tilde {\cal L}}=
\sum_{j=m+1}^{n} {\rm d}p_j\wedge {\rm d}q^j.
\label{twoterms}
\end{equation}
Furthermore the combinations $z^k=q^k+{\rm i}p_k$, $k=1,\ldots, m$, 
are holomorphic coordinates on ${\cal L}$, while the combinations 
$w^j=q^j+ {\rm i} p_j$, $j=m+1, \ldots, n$, are coordinates on $\tilde{\cal L}$. 
In this way a set of coordinates around $b$ is
\begin{equation}
z^1, \ldots, z^{m}, w^{m+1}, \ldots, w^n.
\label{quacoords}
\end{equation}
The holomorphic leaf ${\cal L}$ passing through $b$ may be taken to be
determined by
\begin{equation}
w^{m+1}=0,\ldots, w^n = 0,
\label{pass}
\end{equation}
and spanned by the remaining coordinates $z^k$, $k=1,\ldots, m$.

The construction of the previous sections can be applied as follows. 
Coherent states $|z^k;w^j\rangle$ can be defined locally on 
${\cal C}$. They cannot be extended globally over all of ${\cal C}$, 
as the latter is not a complex manifold. However the foliation 
consists of holomorphic submanifolds. On each one of them there 
exist global coherent states specified by equations 
(\ref{quacoords}), (\ref{pass}), {\it i.e.}, 
\begin{equation}
|z^k;w^{m+1}=0,\ldots, w^n = 0\rangle.
\label{speci}
\end{equation}
Physically, this case corresponds to a fixed splitting of the $n$ degrees 
of freedom in such a way that the first $m$ of them give rise to global coherent 
states on the holomorphic leaves. On the latter there is no room 
for nontrivial dualities. On the contrary, the last $n-m$ degrees of freedom
are only locally holomorphic on ${\cal C}$. Holomorphicity is lost 
globally on ${\cal C}$, thus allowing for the possibility of nontrivial 
duality transformations between different holomorphic leaves. 

Let us analyse the resolution of unity in terms of the states
$|z^k;w^j\rangle$. With the notations of section \ref{prov}, the expansion
\begin{equation}
\sum_{\alpha}\sum_{k=1}^m\sum_{j=m+1}^n\int_{\cal C} {\rm d}\mu_{\cal C}\, 
|z^k_{\alpha};w^j_{\alpha}\rangle  \langle w^j_{\alpha};z^k_{\alpha}|
\label{newres}
\end{equation}
cannot be equated to the identity,
for the same reasons as in section \ref{prov}.
However, integrating over the $w^l$, the expansion
\begin{equation}
\sum_{\alpha}\sum_{k=1}^m\int_{\cal L} {\rm d}\mu_{\cal L}\, 
|z^k_{\alpha}\rangle  \langle z^k_{\alpha}|
\label{newresx}
\end{equation}
can be equated to the identity. The integral extends over any one leaf 
of the foliation. On the contrary, integrating over the $z^k$ in eqn. 
(\ref{newres}) would not give a resolution of the identity.

\subsection{Examples}\label{adesem}

As explained in section \ref{suma}, all physical systems appearing in standard applications 
of quantum mechanics admit a unique complex structure compatible with the symplectic structure 
on their classical phase spaces.  The harmonic oscillator, the 
Coulomb potential, and angular--momentum degrees of freedom, all fall into this category. 
Specifically, the $n$--dimensional harmonic oscillator has ${\cal C}={\bf C}^n$, 
a particle in a Coulomb potential has ${\cal C}=T^*(S^2)\times T^*({\bf R^+})$, 
where $T^*(S^2)$ and $T^*({\bf R^+})$ are the cotangent bundles to the sphere $S^2$ and the half axis, 
repectively, a spherical rotor has ${\cal C}=T^*(S^2)$, and a spin--$1/2$ system (inasmuch as 
it possesses a classical phase space) has ${\cal C}=S^2$. Therefore we cannot expect 
to find examples of nontrivial dualities within standard quantum mechanics. 
In fact this is good news, since we do not wish to modify quantum mechanics as we know it. 
Rather, our goal is that of complementing it, so it will accommodate dualities 
once we go beyond its usual applications.

Outside these well--known systems one can find examples of classical phase spaces 
that meet the criteria of section \ref{nonint}. It suffices to take for 
${\cal C}$ the Cartesian product of a symplectic but {\it non}\/complex space ${\cal C}_1$, 
times a complex space ${\cal C}_2$. For the latter we can consider any of the 
above. An example of the former can be found in ref. \cite{MCDUFF}, where 
${\cal C}_1$ is a nontrivial $U(1)$--bundle over 
another $U(1)$--bundle over the torus $T^2$.  The product of this ${\cal 
C}_1$ with, say, ${\cal C}_2=T^*(S^2)$ (a spherical rotor) admits 
a foliation by holomorphic leaves (${\cal C}_2$), whose complementary leaves 
(${\cal C}_1$) are symplectic but nonholomorphic.

{\bf Acknowledgements}

It is a pleasure to thank J. de Azc\'arraga, U. Bruzzo  and M. Schlichenmaier
for encouragement and technical discussions. This work has been 
partially supported by research grant BFM2002--03681 from Ministerio de 
Ciencia y Tecnolog\'{\i}a and EU FEDER funds.

\end{document}